\documentclass[12pt]{article}
 \def\theequation{\arabic{section}.\arabic{equation}}
\newcounter{rown}
\def\bl{\setcounter{rown}{\value{equation}}
        \stepcounter{rown}\setcounter{equation}0
        \def\theequation{\thesection.\arabic{rown}\alph{equation}}}
\def\el{\setcounter{equation}{\value{rown}}
        \def\theequation{\thesection.\arabic{equation}}}

\usepackage[centertags]{amsmath}
\usepackage{color}
\usepackage{amsfonts}
\usepackage{amssymb}
\usepackage{amsthm}
\usepackage{newlfont}

\begin{document}
\begin{flushright}
FTUV--04-0422  \quad IFIC--04--16 \\April 2004 IFT UWr 1/04
   \\[1cm]
\end{flushright}
\begin{center}

\vspace{.7cm} {\Large {\bf Massive Relativistic Particle Model
with  Spin and Electric Charge from Two-Twistor Dynamics}}

\bigskip
\renewcommand{\thefootnote}{\alph{footnote}}

{\bf Andreas Bette}${}^{+}$\footnote{E-mail: {\tt bette@kth.se}}
 {\bf Jos\'e A. de Azc\'arraga}${}^{\dag}$\footnote{E-mail:
 {\tt j.a.de.azcarraga@ific.uv.es}}, {\bf Jerzy Lukierski}
 ${}^{\ddag}$\footnote{E-mail:
 {\tt lukier@ift.uni.wroc.pl}},\\
 {\bf C. Miquel-Espanya} ${}^{\dag}$\footnote{E-mail:
 {\tt Cesar.Miquel@ific.uv.es}}
 \vskip 0.4truecm

\vskip 0.2cm ${}^{+}$\ {\it Royal Institute of Technology,
 S-151 81 S\"{o}dert\"{a}lje, Sweden}

\vskip 0.2cm ${}^{\dag}$\ {\it Departamento de F\'{\i}sica
Te\'orica and IFIC (Centro Mixto CSIC-UVEG),
 46100-Burjassot (Valencia), Spain}

\vskip 0.2cm ${}^{\ddag}$\ {\it Institute of Theoretical
 Physics, Wroc{\l}aw University, 50-204 Wroc{\l}aw, Poland }

\end{center}

\bigskip

\begin{abstract}
The sixteen real coordinates of two-twistor space are
 transformed by a nonlinear mapping into an enlarged
 space-time framework. The standard relativistic phase
 space of coordinates $(X_\mu, P_\mu)$ is supplemented
 by a six-parameter spin phase manifold (two pairs
 $(\eta_\alpha,\sigma_\alpha)$ and
 $(\overline{\eta}_{\dot\alpha},
 \overline{\sigma}_{\dot{\alpha}})$ of canonically
 conjugated Weyl spinors constrained by two second
 class constraints) and the  electric charge phase
 space ($e,\phi$). The free two-twistor classical
 mechanics is rewritten in this enlarged relativistic
 phase space as a model for a relativistic particle.
 Definite values  for the mass, spin and the electric
 charge of the particle are introduced by means of
 three first class constraints.
\end{abstract}

\newpage
\section{Introduction}

\setcounter{footnote}{0}

Twistor theory (see {\it e.g.}
 \cite{bael1}--\cite{Tod77}) offers an alternative
 geometric picture of space-time physics based on
 the following basic propositions:

\begin{description}
\item{i)} The basic geometry is spinorial and
 conformal {\it i.e.}, the fundamental elementary
 objects are massless.
\item{ii)} Space-time points as well as the
 momentum and other generators of the conformal
 symmetries are composite, and given in terms of
 fundamental conformal spinors -the twistors.
\item{iii)} The appearance of mass, spin  and
 charge of the elementary  object is a result
 of a composite twistor structure; the mass
 parameter breaks the conformal invariance down
 to the Poincar\'{e} one.
\end{description}

One of the important tasks is to translate the
 multitwistor (in particular two-twistor)
 geometry into an extended space-time framework,
 a step that only recently has been completed
 \cite{bael6}--\cite{bael8}. Indeed, in the
 standard Penrose approach (see {\it e.g.}
 \cite{bael1}--\cite{Tod77}) space-time points
 are given by the relation

\begin{align}\label{bael1}
z^{\alpha\dot{\beta}} & =  \frac{i}{f}
 \left( \omega^{{\alpha}}
 \overline{\eta}^{\dot{\beta}} -
 \lambda^{\alpha}\overline{\pi}^{\dot{\beta}}
 \right)\, , & z^{\alpha\dot\beta} &=
 \frac{1}{2}\sigma^{\alpha\dot\beta}_\mu
 z^\mu \, ,
\end{align}
where

\begin{equation}\label{bael2}
f \, = \,
\overline{\pi}^{\dot{\alpha}}
 \overline{\eta}_{\dot{\alpha}}\, ,
\end{equation}
Expression (\ref{bael1}) describes the composite complex
 Minkowski coordinates $z_\mu = x_\mu + iy_\mu$ as a
 solution of the two Penrose incidence equations
 \cite{bael1}--\cite{Tod77}

\begin{equation}\label{bael3}
\omega^{\alpha} = iz^{\alpha
\dot{\beta}}\overline{\pi}_{\dot{\beta}}\, ,\qquad
\lambda^{\alpha} = iz^{\alpha
\dot{\beta}}\overline{\eta}_{\dot{\beta}} \, ,
\end{equation}
involving two twistors $Z^{A}_{i}$
 ($A=1, \ldots, 4;\; i=1,2$) defined by two pairs
 of complex Weyl
spinors,

\begin{equation}\label{bael4}
Z^{A}_{1} = \left(\omega^{\alpha},
 \overline{\pi}_{\dot{\alpha}} \right)\, , \qquad
 Z^{A}_{2} = \left( \lambda^{\alpha},
 \overline{\eta}_{\dot{\alpha}} \right)\, .
\end{equation}

The non-zero fundamental twistorial Poisson brackets (TPB)
 are given by the holomorphic relations
\bl
\begin{eqnarray}\label{bael5a}
\left\{\pi_\alpha , \omega^\beta\right\} =& i
 \delta^{\beta}_{\alpha}\quad  , &\quad
 \left\{\eta_\alpha , \lambda^\beta\right\} =
 i \delta^{\beta}_{\alpha}\, ,\\
 \left\{\overline{\pi}_{\dot\alpha} ,
 \overline{\omega}^{\dot\beta}\right\} =& -i
 \delta^{\dot\beta}_{\dot\alpha}\quad , &
 \quad \left\{\overline{\eta}_{\dot\alpha} ,
 \overline{\lambda}^{\dot\beta}\right\} =
 -i \delta^{\dot\beta}_{\dot\alpha}\, ,\label{bael5b}
\end{eqnarray}
\el
They correspond to the two-twistor symplectic two-form
 ($i=1,2$)

\begin{equation}\label{bael7}
\Omega = d\Theta = i dZ^{A}_{i} \wedge
 d\overline{Z}_{Ai}\, ,
\end{equation}
where $\Theta$ is the Liouville one-form, that may be
 expressed by

\begin{equation}\label{bael6}
\Theta =\frac{i}{2}\left( Z^{A}_{i} d\overline{Z}_{Ai}
 - \overline{Z}^A_{i} dZ_{Ai} \right)\, .
\end{equation}

Now comes a crucial point: using the relations
 (\ref{bael1}), (\ref{bael5a}) and (\ref{bael5b}) one
 can calculate the TPB of the real composite
 Minkowski coordinates $x_\mu = R e z_\mu$. It turns out (see \cite{bael9}) that

\begin{equation}\label{bael8}
\left\{ x_\mu , x_\nu \right\} = - \frac{1}{m^4} \,
\epsilon_{\mu\nu\rho\tau} W^{\rho}P^{\tau}\, ,
\end{equation}
where

\bl
\begin{eqnarray}\label{bael9a}
P^{\mu} & = & \sigma^{\mu}_{\alpha\dot{\beta}}
 \left( \pi^\alpha \overline{\pi}^{\dot{\beta}}
 + \eta^{\alpha}\overline{\eta}^{\dot{\beta}}\right)
 \, ,\\
 \label{bael9b}W^{\mu} & = &\sigma^\mu_{\alpha\dot{\beta}}
 \left[k\, \left( \pi^{\alpha} \overline{\pi}^{\dot{\beta}}
 - \eta^{\alpha} \overline{\eta}^{\dot{\beta}}\right)+
 \rho \eta^{\alpha} \overline{\pi}^{\dot{\beta}} +
 \overline{\rho}\pi^{\alpha}
 \overline{\eta}^{\dot{\beta}}\right]\, ,
\end{eqnarray}
and
\begin{eqnarray}\label{bael9c}
\rho & = &\omega^\alpha\eta_\alpha +
 \overline{\lambda}^{\dot\alpha}\overline{\pi}_{\dot\alpha}\, ,\\
 \label{bael9d}\overline{\rho} &=& \overline{\omega}^{\dot\alpha}
 \overline{\eta}_{\dot\alpha}+\lambda^\alpha\pi_\alpha \, ,\\
 \label{bael9e} k &= & \overline{k}\ =\ \frac{1}{2} \,
 \left(\omega^\alpha\pi_\alpha +\overline{\omega}^{\dot\alpha}
 \overline{\pi}_{\dot\alpha}- \lambda^{\alpha}\eta_\alpha-
 \overline{\lambda}^{\dot\alpha}\overline{\eta}_{\dot\alpha}
 \right)\, .
\end{eqnarray}
\el Thus, the TPB of the space-time coordinates is non-zero,
 {\it i.e.}  after quantization the composite space-time coordinates  will
 become  {\it non-commutative}.

The composite fourvector (\ref{bael9b}) may be identified
 with the Pauli-Luba\'{n}ski vector that describes spin
 in an arbitrary relativistic frame. It is orthogonal,
 as it should, to the momentum (\ref{bael9a}),

\begin{equation}\label{bael10}
P^{\mu}W_{\mu} = 0 \, .
\end{equation}
The composite structure of $W^\mu$ in the primary
 twistor variables is further exhibited by expressing
 $k$, $\rho$ and $\overline{\rho}$ in Eq.
 (\ref{bael9b}) as

\bl
\begin{equation}\label{bael11a}
k= \frac{1}{2} \left( t_{11} - t_{22}\right) \, ,
\qquad \rho = t_{12} \, ,\qquad \overline{\rho}=t_{21} \, ,
\end{equation}
where
\begin{equation}\label{bael11b}
t_{ij} = \tau_{ij}^a \, t^a = Z^{A}_{\ i}\overline{Z}_{Aj}\,
 \quad (a=0,1,2,3;\; i,j=1,2;\; A=1, \dots, 4)\, .
\end{equation}
\el The four isospin Pauli matrices $\tau^a$ describe four
 conformal, $U(2,2)$-invariant scalar products in
 two-twistor space $T\otimes T$. The $t^a$ satisfy  the
 $u(2)=su(2)\oplus u(1)$ twistorial Poisson algebra brackets ($r,s,u=1,2,3$)

\begin{equation}\label{bael12bis}
\left\{ t^r , t^s \right\} = \epsilon^{rsu}\, t^u \, ,
 \qquad \left\{ t^0 , t^r \right\} = 0 \, ,
\end{equation}
as it follows from Eqs. (\ref{bael5a}) and (\ref{bael5b}).

The non-commutativity of the space-time coordinates in
 the presence of nonvanishing spin ($W_\mu \neq 0$)
 can be traced back to earlier considerations
 (see {\it e.g.} \cite{bael10,bael11}). We recall, however, that the usual
 classical and quantum relativistic free fields are
 defined on a classical, commutative space-time. In this paper,
 following \cite{bael6}--\cite{bael8}, we shall show
  how to  replace  the non-commutative  composite coordinates $x_\mu$
 by commutative ones $X_\mu$ (see Sect. 2, (\ref{bale2.9}) and (\ref{bale2.9b})).
 In this way, we obtain a standard relativistic phase space
 with TPB
\begin{equation}\label{bael12}
\left\{X_\mu, X_\nu \right\} = 0 \, , \qquad
 \left\{ P_\mu, P_\nu \right\} = 0 \, ,
 \qquad\left\{P_\mu, X_\nu\right\}= \eta_{\mu\nu} \, ,
\end{equation}
with $X_\mu$ and $P_\mu$ constructed out of twistor primary
 coordinates.

Our aim in this letter is to provide
 the geometric basis for the formulation of two-twistor dynamics
 in terms of more physical variables, describing the extended
 commutative space-time framework.
In Sect. 2 we show that the 16-dimensional
 two-twistor phase space, described by the symplectic
 two-form (\ref{bael7}) (or the TPB's (\ref{bael5a}) and
 (\ref{bael5b})), is mapped bijectively on the relativistic
 phase space $(X_\mu, P_\mu)$ enlarged by an eight-dimensional
 manifold $M_8$ providing the values of the mass, the spin and
 the electric charge. It turns out that this additional manifold
 $M_8$  may also be regarded as a subset of a
 ten-dimensional symplectic vector space spanned by two
 Weyl spinors and two scalars
 $ Y_k = [(\eta_\alpha, \overline{\eta}_{\dot\alpha},
 \sigma_\alpha, \overline{\sigma}_{\dot\alpha},  e , \phi); $
 $\   \alpha,\dot\alpha =1,2;  \, k=1\ldots 10]$
 satisfying two Poincar\'{e}-invariant
 second class constraints $R_1, R_2$,
 the first one depending on the four-momentum. We define the
 symplectic structure on $M_8$ by means of the constrained
 variables $Y_k$ satisfying the appropriate Dirac brackets.

In Sect. 3 we introduce an action for a relativistic particle
 model which is inspired by the two-twistor conformal-invariant
  Liouville one-form (\ref{bael6}). The
 eighteen-dimensional space $(X_\mu, P_\mu, Y_k)$ is restricted
 by the two second class constraints $R_1=R_2=0$  as well as
 by three additional first class constraints, corresponding
 after first quantization to the wave equations describing
 mass, spin and electric charge.
 It appears that in such a Lagrangian formulation we reproduce the twistor
 Poisson brackets for composite coordinates $(X_\mu , P_\mu)$ as the Dirac
  brackets.\footnote{We identify TPB and Dirac brackets for particular
  parametrization of $M_8$, given by real projective spinors
  $\widetilde{\eta}_{\alpha}
   = \frac{\eta_{\alpha}}{\overline{\xi}_{\dot\beta} \overline{\eta}^{\dot\beta}},
    \widetilde{\sigma}_{\alpha}=
   \frac{\sigma_{\alpha}} {\overline{\xi}_{\dot{\beta}}
     \overline{\sigma}^{\dot{\beta}}}$  (see
 \cite{bael6,bael7}).}

 We would like to point
 out that in our formulation  all eight relativistic phase space
 coordinates $(X_\mu,P_\mu)$ satisfying the TPB relations
 (\ref{bael12}) are two-twistor composites. This last remark is
 relevant in order to distinguish the present approach from
 earlier attempts to describe the spin degrees of freedom of
 free massive relativistic particles in terms of spinorial
 and twistorial coordinates (see {\it e.g.}  \cite{bael12}--\cite{bael14}),
 in which the space-time manifold was introduced as a
 primary geometric object.

The framework presented in this paper may be considered as
 a generic one. Indeed, it can be further extended by
 adding supersymmetries (see {\it e.g.} \cite{bael15,bael16})
 as well as by going to higher $(D>4)$ dimensions.
 In particular, we would like to notice here that the
 $D=11$ BPS preons, introduced in \cite{bael17} as fundamental
 constituents of BPS states in M-theory,  may be described in
 terms of $D=11$ generalized supertwistors. Finally,
 two-twistor space provides a natural framework  for the
 introduction of  infinite-dimensional higher spin multiplets
 (see \cite{bael18}), with arbitrary masses and charges.

\section{From two-twistor phase space to enlarged
 relativistic phase space.}

\setcounter{equation}{0}

In terms of the 16 components of the two twistors $Z_1, Z_2$,
 and their complex conjugates, the Liouville one-form
 $\Theta$ in Eq. (\ref{bael6}) reads

\begin{equation}\label{bale2.1}
\Theta = \frac{i}{2} \left( \omega^\alpha  d \pi _\alpha
 +\overline{\pi}_{\dot{\alpha}}
 d\overline{\omega}^{\dot{\alpha}} - c.c. \right)+\,
 \frac{i}{2}\left( \lambda^\alpha  d \eta _\alpha +
 \overline{\eta}_{\dot{\alpha}}
 d\overline{\lambda}^{\dot{\alpha}} - c.c.\right) \, .
\end{equation}
Using Eq. (\ref{bael3}) and writing

\begin{equation}\label{bale2.2.}
z^{\alpha \dot{\beta}} = x^{\alpha \dot{\beta}} +
 i y^{\alpha \dot{\beta}} \qquad
 (z^\mu = x^\mu + iy^\mu)\, ,
\end{equation}
one gets:

\begin{eqnarray}\label{bale2.3}
\Theta & = & \pi_\alpha \overline{\pi}_{\dot{\beta}}
 dx^{\alpha \dot{\beta}} + i y^{\alpha \dot{\beta}}
 \left( \pi_\alpha d\overline{\pi}_{\dot{\beta}} -
 \overline{\pi}_{\dot{\beta}} d \pi_\alpha \right)
 \cr \cr && + \eta_\alpha \overline{\eta}_{\dot{\beta}}
 dx^{\alpha \dot{\beta}} + i y^{\alpha \dot{\beta}}
 \left( \eta_\alpha d\overline{\eta}_{\dot{\beta}} -
 \overline{\eta}_{\dot{\beta}} d \eta_\alpha \right)\, .
\end{eqnarray}
Using the definition (\ref{bael9a}) of $P_\mu$ we obtain

\begin{equation}\label{bale2.4}
\Theta = P_{\mu} dx^\mu  + iy^{\alpha \dot{\beta}}
 \left( \pi_\alpha d\overline{\pi}_{\dot{\beta}}
 - \overline{\pi}_{\dot{\beta}} d \pi_\alpha +
 \eta_\alpha d\overline{\eta}_{\dot{\beta}} -
 \overline{\eta}_{\dot{\beta}} d \eta_\alpha \right)\, .
\end{equation}

\medskip
\noindent Further, it can be shown that (see (\ref{bael11a})--(\ref{bael11b}))

\begin{equation}\label{bale2.5}
t_{11}  = - 2y ^{\alpha \dot{\beta}} \pi_\alpha
 \overline{\pi}_{\dot{\beta}} \, , \quad t_{22}  =
 - 2y^{\alpha \dot{\beta}} \eta_\alpha
 \overline{\eta}_{\dot{\beta}}, \quad \rho = t_{12}
 =\overline{t}_{21} = - 2 y^{\alpha \dot{\beta}}
 \eta_\alpha \overline{\pi}_{\dot{\beta}}\, ,
\end{equation}
 where

\begin{eqnarray}\label{bale2.6}
y ^{\alpha \dot{\beta}}& = & \frac{1}{2f\overline{f}}
 \left( \rho \pi^{\alpha} \overline{\eta}^{\dot{\beta}}
 + \overline{\rho} \eta^{\alpha} \overline{\pi}^{\dot{\beta}}
 - t_{11} \eta^{\alpha} \overline{\eta}^{\dot{\beta}} -
 t_{22} \pi^{\alpha} \overline{\pi}^{\dot{\beta}} \right)\, ,
\end{eqnarray}
and $f$ is given by (\ref{bael2}). Subsequently,

\begin{equation}\label{bale2.7}
y^{\alpha \dot{\beta}} \pi_\alpha = \frac{1}{2f}
 \left( t_{11} \overline{\eta}^{\dot{\beta}} - \overline{\rho}\,
 \overline{\pi}^{\dot{\beta}} \right) \, , \qquad
 y^{\alpha \dot{\beta}} \eta_\alpha = \frac{1}{2f} \left(
 \rho \overline{\eta}^{\dot{\beta}} - t_{22}
 \overline{\pi}^{\dot{\beta}} \right)\, ,
\end{equation}
and one gets

\begin{equation}\label{bale2.8}
\Theta  = P_\mu dx^\mu + \left\{ \frac{i}{2f} \left( t_{11}
 \overline{\eta}^{\dot{\alpha}} - \overline{\rho}\,
 \overline{\pi}^{\dot{\alpha}} \right)
 d\overline{\pi}_{\dot{\alpha}} + \frac{i}{2f} \left(
 \rho \overline{\eta}^{\dot{\alpha}} - t_{22}
 \overline{\pi}^{\dot{\alpha}} \right)
 d\overline{\eta}_{\dot{\alpha}} + c.c. \right\}\, .
\end{equation}

We see that the composite space-time  coordinates $x_\mu$,
 which have nonvanishing TPB among themselves given by
 (\ref{bael8}), enter in the one-form
 (\ref{bale2.8}). Thus, we have arrived at a symplectic formalism
 with non-commutative space-time coordinates. In principle, one
 could pursuit the construction of dynamical models of
 non-commutative classical mechanics based on the Liouville
 one-form (\ref{bale2.8}). However, one can do better:
 following recent results \cite{bael6}--\cite{bael8} one can
 define the commutative composite space-time coordinates
 $X_\mu$ by making a shift,
 $x_\mu \to X_\mu = x_\mu + \Delta x_\mu$, in such a way that
 the composite relativistic phase space variables
 $(X_\mu, P_\mu)$ satisfy the standard  relations
 (\ref{bael12}). For that purpose we redefine
 ({\it cf.} Eq.(\ref{bael1}))

\bl
\begin{eqnarray}\label{bale2.9}
z^{\alpha \dot{\beta}} \to Z^{\alpha \dot{\beta}} & =
 & z^{\alpha \dot{\beta}} + \Delta \, z^{\alpha \dot{\beta}}=
 z^{\alpha \dot{\beta}} + i \rho\frac{\pi^{\alpha}
 \overline{\eta}^{\dot{\beta}}}{f\overline{f}} =
 X^{\alpha \dot{\beta}} + i Y^{\alpha \dot{\beta}}\, ,
\end{eqnarray}
where $X^{\alpha\dot\beta}=\frac{1}{2}
 \sigma^{\alpha\dot\beta}_\mu X^\mu$ and
 $X^\mu=\mathrm{Re} Z^\mu = \mathrm{Re} (z^\mu+\Delta z^\mu)$
  or, explicitly,
\begin{eqnarray}
 \label{bale2.9b}
 X^{\mu} &= &
 x_{\mu} + \frac{i}{2f\overline{f}}(\sigma^{\mu})_{\alpha
  \dot{\beta}}
  \left(
  \rho \, \pi^{\alpha} \ \overline{\eta}^{\dot{\beta}}
  - \overline{\rho} \, \eta^{\alpha} \, \overline{\pi}^{\dot{\beta}}
  \right)\, .
\end{eqnarray}
Using the TPB (\ref{bael5a}) and (\ref{bael5b}) one can show that the real coordinates
(\ref{bale2.9b}) commute. \el Further we get

\begin{eqnarray}\label{bale2.10}
P_\mu d(\Delta x^\mu) & = & P_\mu \left( dX^\mu - dx^\mu \right) =
 \cr\cr &=&  \frac{1}{2} \left( \pi_\alpha
 \overline{\pi}_{\dot{\beta}} + \eta_\alpha
 \overline{\eta}_{\dot{\beta}} \right) d \left(
 \frac{i \rho \pi^\alpha \overline{\eta}^{\dot{\beta}}}
 {f\overline{f}} + c.c. \right)\, ,
\end{eqnarray}
and we arrive at the formula

\begin{equation}\label{bale2.11}
\Theta = P_\mu dX^\mu + \left[ \frac{i}{f} \left(
 \rho\overline{\eta}^{\dot{\alpha}} + \,  k
 \overline{\pi}^{\dot{\alpha}} \right)
 d\overline{\eta}_{\dot{\alpha}} + c.c. \right] +
 \frac{i}{2} t_{11} \left( \frac{d\overline{f}}
 {\overline{f}} - \frac{df}{f} \right)\, .
\end{equation}
Introducing

\bl
\begin{eqnarray}\label{bale2.12a}
\sigma^{\alpha} &=& - \frac{1}{\overline{f}} \left(
 \overline{\rho} {\eta}^{{\alpha}} + \, k {\pi}^{{\alpha}}\right)\,
 , \qquad \overline{\sigma}^{\dot{\alpha}} = - \frac{1}{f}
 \left( \rho \overline{\eta}^{\dot{\alpha}} + \,
 k \overline{\pi}^{\dot{\alpha}}\right) \, ,\\
 \cr e & = &  t_{11} \, , \qquad \qquad\qquad \qquad
 \phi  = \frac{i}{2} \ln\frac{\overline{f}}{f}\,
 , \label{bale2.12b}
\end{eqnarray}
\el
one obtains finally

\begin{eqnarray}\label{bale2.13}
\Theta = P_\mu d X^\mu -i \left(
 \overline{\sigma}^{\dot{\alpha}} d\overline{\eta}_{\dot{\alpha}}
 - c.c. \right) +  e d \phi \, .
\end{eqnarray}

The $18$ variables ($X_\mu$, $P_\mu$, $\eta_\alpha$,
 $\overline{\eta}_{\dot{\alpha}}$,  $\sigma_\alpha$,
 $\overline{\sigma}_{\dot{\alpha}}$, $e$, $\phi$)
 are not all independent. Comparing the number of
 degrees of freedom (16 in Eq. (\ref{bale2.1})
 versus 18 in Eq. (\ref{bale2.13})) one can deduce that the
 variables occurring in Eq. (\ref{bale2.13}) satisfy two
 constraints. The first one takes the form
 ($\mathbf{t}^2 = t_1^2 + t^2_2 + t^2_3$):

\begin{eqnarray}\label{bale2.15bis}
R_1 &=& \sigma_\alpha \, P^{\alpha \dot{\beta}}
 \overline{\sigma}_{\dot\beta} -  \mathbf{t}^2 = 0,\quad
 \mathbf{t}^2 = |\rho|^2 + k^2\, .
\end{eqnarray}

In order to show explicitly how Eq. (\ref{bale2.15bis})
 restricts the eighteen variables of our generalized
 phase space ($X_\mu, P_\mu, \eta_\alpha,
 \overline{\eta}_{\dot{\alpha}}, \sigma_\alpha,
 \overline{\sigma}_{\dot{\alpha}},  e, \phi$)
 we observe that

\begin{equation}\label{bale2.16bis}
k =  \overline{\eta}^{\dot{\alpha}}\,
 \overline{\sigma}_{\dot{\alpha}} \, =
 \, {\eta}^{{\alpha}}\,{\sigma}_{{\alpha}}
 = \overline{k}\, ,
\end{equation}
\bl
\begin{eqnarray}\label{bale2.17bis}
\rho  = \overline{\pi}_{\dot{\alpha}}
 \overline{\sigma}^{\dot{\alpha}} = -\frac{1}{\overline{f}}
 \eta_{\alpha} P^{\alpha\dot{\beta}}\overline{\sigma}_{\dot{\beta}}
 \, , & & \overline{\rho} =\pi_{\alpha} \sigma^{\alpha}=
 -\frac{1}{f} \sigma_{\alpha} \, \, P^{\alpha\dot{\beta}}
 {\overline{\eta}}_{\dot{\beta}}\, ,\\
 |\rho|^2 &=& \frac{2}{P^2} P^{\alpha\dot{\beta}}
 P^{\gamma\dot{\delta}} \, {\sigma}_{{\alpha}}\,
 \overline{\eta}_{\dot{\beta}}\,
 {\overline{\sigma}}_{{\dot\delta}}\, {\eta}_{{\gamma}}\, ,
\end{eqnarray}
\el\noindent where
\begin{equation}\label{bale2.22bi}
\pi_\alpha = - \frac{1}{f} \, P_{\alpha \dot{\beta}} \,
 \overline{\eta}^{ \dot{\beta}}\, , \qquad f =
 \frac{1}{\sqrt{2}}\,{P} \ e^{i \phi} \, , \qquad
 P\equiv(P^\mu P_\mu)^{\frac{1}{2}}\, .
\end{equation}

Taking into account Eq. (\ref{bale2.16bis}), the reality
 of the variable $k$ produces the second constraint equation

\begin{eqnarray}\label{bale2.18bis}
R_2 & = &  \eta_{\alpha}\, \sigma^{\alpha} -
 \overline{\eta}_{\dot{\alpha}}\,
 \overline{\sigma}^{\dot{\alpha}} = 0 \, .
\end{eqnarray}
We point out that if we employ the composite formulae
 (\ref{bael9a}) and (\ref{bale2.12a}) both $R_1$ and
 $R_2$ are identically zero in terms of the two-twistor
 variables (\ref{bael4}).

If we select as generalized momenta the set of the nine
 commuting variables \sloppy $(P_\mu, \eta_\alpha,
 \overline{\eta}_{\dot{\alpha}}, e)$
 they satisfy also  the following constraint
 ($P^2 \equiv P_{\alpha\dot{\beta}}
 P^{\alpha \dot{\beta}}=2f\overline{f}$)

\begin{eqnarray}\label{bale2.14}
\eta_\alpha P^{\alpha\dot{\beta}}
 \overline{\eta}_{\dot{\beta}}& = & \frac{1}{2}\, P^2\, .
\end{eqnarray}
It can be shown, however, that the constraint (\ref{bale2.14})
 follows from the constraints (\ref{bale2.15bis}) and
 (\ref{bale2.18bis}). Therefore,  using (\ref{bale2.14}) the mass
 $m$ can be defined by the following subsidiary condition:

\begin{eqnarray}\label{bale2.19bb}
R_3 & = & \eta_{\alpha} P^{\alpha \dot{\beta}}
 \overline{\eta}_{\dot{\beta}} - \frac{1}{2} m^2 = 0 \, .
\end{eqnarray}

Much in the same way as the constraint (\ref{bale2.14}) provides
 the mass Casimir $P^2$, the constraint (\ref{bale2.15bis})
 defines the square of the relativistic spin operator.
 Its numerical value provides the fourth constraint:

\begin{eqnarray}\label{bale2.20bb}
R_4 & = & \sigma_{\alpha} \, P^{\alpha \dot{\beta}}
 \overline{\sigma}_{\dot{\beta}}\, - s(s+1) =0 \, .
\end{eqnarray}

\noindent
Indeed, from the formula (\ref{bael9b}) defining the
 Pauli-Luba\'{n}ski relativistic spin fourvector
 $W^\mu$ it follows that

\begin{eqnarray}\label{bale2.19bis}
&&k  = \frac{1}{2|f|^2}\left(
 \eta_{\alpha}\overline{\eta}_{\dot{\beta}} -
 \pi_{\alpha} \, \overline{\pi}_{\dot{\beta}}
 \right) W^{\alpha\dot{\beta}} \, ,
 \cr\cr  && \rho = - \frac{1}{|f|^2}\ \pi_{\alpha}
 \, \overline{\eta}_{\dot{\beta}}W^{\alpha\dot{\beta}}\,
 ,\qquad \overline{\rho} = - \frac{1}{|f|^2} \
 \eta_{\alpha} \, \overline{\pi}_{\dot{\beta}}
 W^{\alpha\dot{\beta}}\, .
\end{eqnarray}
Using Eq. (\ref{bael9b}),
one obtains ($W^2 \equiv W_{\alpha\dot{\beta}}
 W^{\alpha\dot{\beta}}=  W_\mu W^\mu$):

\begin{eqnarray}\label{bale2.20bis}
\mathbf{t}^2 & = & - \frac{1}{2f\overline{f}} \,
 W_{\alpha\dot{\beta}} W^{\alpha\dot{\beta}} = -
 \frac{1}{P^2} \, W^2 \, .
\end{eqnarray}
Because  $\{P_\mu, \rho\}=\{P_\mu , k\}=0$ one
 gets $\{P_\mu, \sigma_\alpha\}=0$ and (Eqs.
 (\ref{bale2.19bb}), (\ref{bale2.20bb}))

\begin{eqnarray}\label{bale2.21bis}
\left\{ \mathbf{t}^2, P^2 \right\} & = & 0 \, .
\end{eqnarray}

We see therefore that the constraints $R_3=R_4=0$ on
 the constrained 16-dimensional generalized phase space
 take the form ($m\geq 0; s=0,\frac{1}{2}, 1, \ldots$),
 on account of Eqs. (\ref{bale2.14}) and (\ref{bale2.15bis}),

\begin{eqnarray}\label{bale2.22bis}
P^2 = m^2 \, , \qquad \mathbf{t}^2 = s(s+1) \, ,
\end{eqnarray}

\noindent
and provide the mass and spin values characterizing a
 relativistic particle. The electric charge $e_0$ is
 defined by the fifth subsidiary condition
\begin{equation}\label{bale2.25} R_5 = e - e_0 = 0 \, ,
\end{equation}

\noindent
where $e$ is given by (\ref{bale2.12b}).

\section{Relativistic particle model from  the two-twistor framework}

\setcounter{equation}{0}

Looking at the Liouville one-form (\ref{bale2.13}),
 with all 18 coordinates $X_\mu,P_\mu,Y_k$ now treated
 as primary, we propose the following action for a charged,
 massive relativistic particle with spin

\begin{eqnarray}\label{bale3.1}
S   & = &\int d\tau\mathcal{L} = \int d\tau \left[ P_\mu \dot{X}^\mu +
 i\,( \sigma^{\alpha} \dot{\eta}_{\alpha} -
 \overline{\sigma}^{\dot{\alpha}}
 \dot{\overline{\eta}}_{\dot{\alpha}})+
 e \dot{\phi} \right. \cr\cr
 && \left.+ \lambda_1 R_1 +
 \lambda_2 R_2 + \xi_1(P^2 -m^2)+
 \xi_2(\mathbf{t}^2 - s(s+1)) +\xi_3(e-e_0)\right]\, ,\nonumber\\
\end{eqnarray}
where {\it e.g.} $\dot{\overline{\eta}}_{\dot\alpha}=
 d\overline{\eta}_{\dot\alpha}/d\tau$ and the $\lambda$'s
 and $\xi$'s are Lagrange multipliers.

The canonical Poisson brackets determined by the
 action (\ref{bale3.1}) will be denoted by
 $\{\, \cdot\,  , \, \cdot \, \}_{C}$. They are

\bl
\begin{eqnarray}\label{bale3.2a}
&& \quad \left\{ X_\mu , X_\nu \right\}_{C} = 0 \,
 ,\quad \left\{ P_\mu , P_\nu \right\}_{C} = 0 \,
 ,\quad \left\{ P_\mu , X_\nu \right\}_{C} = \eta_{\mu\nu}\, ,
 \\ \cr
 \label{bale3.2b} && \qquad \quad \left\{ \eta_{\alpha} ,
 \sigma^{\beta} \right\}_{C} =i\delta_{\alpha}^{\ \beta}
 \, , \qquad \left\{ \overline{\eta}_{\dot{\alpha}} ,
 \overline{\sigma}^{\dot{\beta}}\right\}_{C} = -i
 \delta_{\dot{\alpha}}^{\ \dot{\beta}} \, ,\\
 \cr \label{bale3.2c} && \qquad \qquad \qquad
 \qquad \qquad \left\{ e, \phi\right\}_{C} = 1 \, ,
\end{eqnarray}
\el all others being zero. The five constraints described
 by the action (\ref{bale3.1}) via the Lagrange multipliers
 split into a pair, (\ref{bale2.15bis}) and (\ref{bale2.18bis}),
 of second class constraints, reducing the number of degrees
 of freedom from 18 to 16, and three first class constraints,
 Eqs. (\ref{bale2.19bb})--(\ref{bale2.20bb}) (or equivalently
 (\ref{bale2.22bis})) plus Eq. (\ref{bale2.25}). These three
 first class constraints define the mass, spin and electric charge
 values, which further reduce the number of degrees of freedom
 from 16 to 10.

Let us observe that in order to obtain the consistency of the
 Poisson structure with the second class constraints
 $R_1 =R_2=0$, the canonical PB
 \hbox{(\ref{bale3.2a})--(\ref{bale3.2c})} have to be replaced
 by Dirac brackets given by

\begin{eqnarray}\label{bael12b}
\left\{ Y_k, Y_l\right\}_{ D} &=& \left\{
 Y_k, Y_l \right\}_{C} + \left\{ Y_k, R_1 \right\}_{C}
 {\frac{1}{\{ R_1, R_2\}}}_{C} \left\{ R_2, Y_l \right\}_{C}
 \cr\cr && - \left\{ Y_k, R_2 \right\}_{C}
 {\frac{1}{\{ R_1, R_2\}}}_{C} \left\{ R_1, Y_l \right\}_{C}\, ,
 \end{eqnarray}
where\\
\begin{equation}\label{bale3.4}
\left\{ R_1 , R_2 \right\}_C = -2i \sigma_\alpha \,
 P^{\alpha \dot{\beta}}\overline{\sigma}_{\dot{\beta}} \, .
\end{equation}

  For the spin sector variables
 $(\eta_\alpha, {\overline{\eta}}_{\dot{\alpha}},
 \sigma_\alpha, \overline{\sigma} _{\dot{\alpha}})$
 one gets
\bl
\begin{eqnarray}\label{bale3.5a}
\left\{ \eta_{\alpha} , \eta_{\beta} \right\}_{D} & =
 & \left\{ \overline{\eta}_{\dot{\alpha}} ,
 \overline{\eta}_{\dot{\beta}} \right\}_{D} =
 \left\{ {\eta}_{{\alpha}} , \overline{\eta}_{\dot{\beta}}
 \right\}_{D} = 0 \, ,\\
 \cr \left\{ \sigma^{\alpha} , \sigma^{\beta} \right\}_{D} & =
 &\frac{-i}{P^2}(P^{\alpha\dot\gamma}
 \overline{\eta}_{\dot\gamma}\sigma^\beta- P^{\beta\dot\gamma}
 \overline{\eta}_{\dot\gamma}\sigma^\alpha)\, ,\\
 \cr \left\{ {\sigma}^{{\alpha}} , \overline{\sigma}^{\dot{\beta}}
 \right\}_{D}& =&\frac{-i}{P^2}(P^{\alpha\dot\gamma}
 \overline{\eta}_{\dot\gamma}\overline{\sigma}^{\dot\beta}+
 \eta_\gamma P^{\gamma\dot\beta}\sigma^\alpha)\, ,\label{bale3.5cc}\\
 \cr \left\{ \overline{\sigma}^{\dot{\alpha}} ,
 \overline{\sigma}^{\dot{\beta}} \right\}_{D} & =
 &\frac{i}{P^2}(\eta_\gamma P^{\gamma\dot\alpha}
 \overline{\sigma}^{\dot\beta}-\eta_\gamma
 P^{\gamma\dot\beta}\overline{\sigma}^{\dot\alpha})\, ,\\
 \cr \left\{{\eta}_{\alpha} , \sigma^{\beta}
 \right\}_{D} & = & i \delta_{\alpha}^{\ \beta} -
 \frac{i}{P^2} \eta_\alpha P^{\beta \dot{\gamma}}
 \overline{\eta}_{\dot{\gamma}} \, ,\\
 \cr\left\{\overline{\eta}_{\dot{\alpha}} ,
 \overline{\sigma}^{\dot{\beta}} \right\}_{D}
 &= & -i\,\delta_{\dot\alpha}^{\dot\beta} +
 \frac{i}{P^2} \eta_\gamma P^{\gamma \dot{\beta}}
 \overline{\eta}_{\dot{\alpha}} \, ,\\
 \label{bale3.5e}\cr \left\{ \eta_{\alpha} ,
 \overline{\sigma}^{\dot{\beta}} \right\}_{D} & = &
 \frac{i}{P^2} \eta_\alpha \eta_\gamma P^{\gamma \dot{\beta}}\, .
\end{eqnarray}
\el
The relations
 (\ref{bale3.5a})--(\ref{bale3.5e}) provide the Dirac
 bracket structure for the six degrees of freedom of the spin
 phase space $(\eta_\alpha, \overline{\eta}_{\dot{\alpha}},
 \sigma_{\alpha}, \overline{\sigma}_{\dot{\alpha}})$,
  consistent with  the constraints  $R_1=0=R_2$.
 Further, we have
\begin{equation}\label{bale3.6}
\left\{\eta_{{\alpha}} , P^{\beta\dot{\gamma}} \right\}_{D} =
 \left\{\overline{\eta}_{{\dot{\alpha}}} , P^{\beta\dot{\gamma}}
 \right\}_{D} =\left\{\sigma_{{\alpha}} , P^{\beta\dot{\gamma}}
 \right\}_{D}=\left\{\overline{\sigma}_{\dot{\alpha}} ,
 P^{\beta\dot{\gamma}}\right\}_{D} = 0 \, ,
\end{equation}
but
\begin{eqnarray}\label{bale3.7}
\left\{\eta_{{\alpha}} , X^{\beta\dot{\gamma}}
 \right\}_{D} &= & \eta_{\alpha}
 \frac{\widetilde{P}^{\beta\dot{\gamma}}}{P^2} \, ,
\qquad \left\{\sigma_{{\alpha}} ,
 X^{\beta\dot{\gamma}} \right\}_{D} =
 -\sigma_{\alpha}\frac{\widetilde{P}^{\beta\dot{\gamma}}}{P^2} \, ,
\end{eqnarray}
where
\begin{eqnarray}\label{bale3.8bi}
\widetilde{P}^{\beta \dot{\gamma}}&=& P^{\beta\dot\gamma}
 \frac{k^2}{\rho\overline{\rho}+k^2} -
 \eta^\beta\overline{\eta}^{\dot\gamma}\, .
\end{eqnarray}
Because $\phi$ and $e$ have vanishing Poisson brackets
 with $R_1$ and $R_2$ one obtains

\begin{equation}\label{bale3.8}
\left\{e,\phi\right\}_{D} = 1\, .
\end{equation}
Finally, since $\{R_{1,2},P\}_C=0=\{R_2, X^{\alpha\dot\beta}\}_C$,
 the Dirac brackets in the relativistic phase sector
 $(X_\mu, P_\nu)$ coincide with the canonical ones given by Eq.
 \hbox{(\ref{bale3.2a})}.

   The canonical Poisson
 brackets (\ref{bale3.2a}) are  useful if we calculate
 from the action (\ref{bale3.1}) the Noether charge
 $\Sigma_{\mu\nu}$, describing the generators of the Lorentz
 transformations in the spin sector {\it i.e.}, the spin part
 of the relativistic angular momentum. One obtains, using spinorial
 notation (see also \cite{bael8})
\begin{equation}\label{bale3.10}
\Sigma_{\mu\nu} = \frac{1}{2i}\sigma_\mu^{\alpha\dot\alpha}
 \sigma_\nu^{\beta\dot\beta} (\sigma_{(\alpha}\ \eta_{\beta)} \
 \epsilon_{\dot{\alpha}\dot{\beta}} -
 \overline{\sigma}_{(\dot{\alpha}}\
 \overline{\eta}_{\dot{\beta})} \epsilon_{\alpha \beta})=
 -\Sigma_{\nu\mu}\, ,
\end{equation}
where the symmetrization is with unit weight. From
 (\ref{bale3.2b}) follows that\footnote{Using that the
 total generators of the Lorentz transformations
 $M^{\mu\nu}=2P^{[\mu}X^{\nu]}+\Sigma^{\mu\nu}$
 fulfil the Lorentz algebra and that they have vanishing
 PB with the two second class constraints $R_1,R_2$, one can also show that
 $\{\Sigma_{\mu\nu},\Sigma_{\rho\tau}\}_D=
 \{\Sigma_{\mu\nu},\Sigma_{\rho\tau}\}_C\quad .$}

\begin{eqnarray}\label{bale3.11}
\left\{ \Sigma_{\mu\nu}, \Sigma_{\rho\tau} \right\}_{C}
 = \eta_{\mu\tau}\Sigma_{\rho\nu} - \eta_{\mu\rho}
 \Sigma_{\nu\tau} + \eta_{\rho\nu} \Sigma_{\mu\tau} -
 \eta_{\nu\tau} \Sigma_{\mu\rho}
\, .
\end{eqnarray}
It is easy to show that the Lorentz spin generators
 (\ref{bale3.10}) imply the following values of the
 two Lorentz Casimirs:
\bl
\begin{eqnarray}\label{bale3.12a}
C_1 & = & \frac{1}{2}\,\Sigma_{\mu\nu} \Sigma^{\mu\nu}
 = \frac{1}{2} \left[ (\eta_{\alpha} \,
 \sigma^{\alpha})^2 + ( \overline{\eta}_{\dot{\alpha}}
 \, \overline{\sigma}^{\dot{\alpha}})^2 \right] =
 k^2\, , \\
 C_2 & = & \frac{1}{8}\, \Sigma_{\mu\nu}
 \epsilon^{\mu\nu\rho\tau} \Sigma_{\rho\tau} = 0 \,
 ,\label{bale3.12b}
\end{eqnarray}
\el where Eq. (\ref{bale2.16bis}) has been used. We notice
 that $C_2\neq 0$ requires having $n$-twistor coordinates
 with $n>2$.

In order to quantize the model described by the action
 (\ref{bale3.1}), the quantum counterparts of the
 first class constraints $R_3=R_4 = R_5 = 0$ are
 needed. One can proceed in two different ways.

i) The covariant formulation consists in  treating
 the three first class constraints as conditions on
 quantum states which lead to wave equations. For this
 aim one  considers the differential realization of the
 Dirac brackets (\ref{bale3.5a})--(\ref{bale3.7}) on the
 generalized momentum space ($P_\mu, \eta_\alpha,
 \overline{\eta}_{\dot{\alpha}}, \phi$) that provides
 the Schr\"{o}dinger representation for the
 first-quantized theory.

ii) Another way is to look at the first class constraints
 $R_3 = R_4= R_5 =0$ as generators of three local symmetries,
 and to fix the gauge of the corresponding local degrees
 of freedom. In such a way the three gauge-fixing conditions
 plus the constraints $R_3=R_4=R_5=0$ provide
 six additional second  class constraints. These reduce the
 $16$ degrees of freedom of two-twistor space to the
 ``physical'' ten-dimensional generalized phase space.
 In this formulation the gauge-fixing conditions break necessarily
 the Lorentz invariance and lead to a noncovariant formulation
 of the first-quantized theory in the Heisenberg picture.

Both methods of quantization are under consideration by
 the present authors.

\section{Conclusions}

\protect\hspace{12pt} The main aim of this paper was to
 show how to move from the two-twistor geometry to a
 generalized space-time description of relativistic
 particle mechanics. In order to introduce our particle
 model action, Eq. (\ref{bale3.1}), we used the
 symplectic potential (Liouville one-form) (\ref{bale2.13})
 obtained from the composite nature of the variables
 $(X_\mu, P_\mu, \sigma_\alpha, \overline{\sigma}_{\dot{\alpha}},
 e, \phi)$; only  $\eta_\alpha, \overline{\eta}_{\dot{\alpha}}$
 remain as  primary twistor coordinates. We stress again that
 in such a framework both the fourmomentum $P_\mu$ (Eq.
 (\ref{bael9a})) as well as the real Minkowski space-time
 coordinates $X_\mu$ (Eq. (\ref{bale2.9b})) are composite.
 Nevertheless, in the action (\ref{bale3.1}) all the eighteen
 variables $X_\mu,P_\mu,Y_k$ $(k=1,\ldots, 10)$ are taken as
 primary ones and only the presence of the constraints
 $R_1 = R_2=0$ exhibits their twistorial  origin. On
 our sixteen-dimensional   generalized phase space
 $\{X_\mu, P_\mu, M_8  \}$ one can introduce
 two Poisson structures
 consistent with the pair of second class constraints:
 one  $\{ \, \cdot \, , \, \cdot \, \}$,
 induced by the fundamental TPB (\ref{bael5a}) and (\ref{bael5b})
 and applied to the composite twistor formulae, and a second one
 $\{ \, \cdot \, , \, \cdot \, \}_{D}$, given by the
 Dirac brackets (\ref{bael12b}) and further explicitly
  calculated in (\ref{bale3.5a})--(\ref{bale3.8}).

  We would like to report here an important calculational result:  for
  the 8-dimensional parametrization of $M_8$, given by the
  projective real spinors $\widetilde{\eta}_{\dot{\alpha}} =
  \frac{\eta_{\alpha}}{\overline{\xi}_{\dot\alpha}
  \overline{\eta}^{\dot{\alpha}}}$,
  $\widetilde{\sigma}_{{\alpha}} =
  \frac{\sigma_{\alpha}}{\overline{\xi}_{\dot\alpha}
  \overline{\sigma}^{\dot{\alpha}}}$ ($\xi_\alpha$ is a  constant
  spinor; see \cite{bael6,bael7}) supplemented by the pair
 of variables $(e,\varphi)$, these two Poisson
   structures are the same.

In this paper we have limited ourselves to the classical
 theory and provided an outline on how to construct the
 first-quantized theory. Clearly, the first-quantized
 theory is important because it provides the description
 of a class of relativistic free fields, those with spin
 generators restricted by the constraint (\ref{bale3.12b}).
 Our final aim  is to obtain the covariant description of
 all massive Wigner representations of the
 Poincar\'{e} group as solutions of the first-quantized
 free particle model (\ref{bale3.1}), or of its generalization
 to a three-twistor space. Finally, we recall that our
 two-twistor framework also provides, besides the spin
 description, the canonical pair of variables
 (see (\ref{bale3.8})) describing a $U(1)$ internal gauge
 degree of freedom as well as the  electric charge.

\bigskip

\noindent {\it Acknowledgements}. This work has been partially
 supported by the Spanish Ministerio de Educaci\'{o}n y
 Ciencia through grant BFM2002-03681 and EU FEDER funds.
 One of us (C.M.E.) wishes to thank the Spanish M.E.C.
 for his research grant.

\end{document}